\documentclass[%
 reprint,
 amsmath,amssymb,
 aps,
 floatfix,
 superscriptaddress
]{revtex4-2}

\usepackage{graphicx}
\usepackage{dcolumn}
\usepackage{bm}
\usepackage{float}

\begin{document}

\title{Anomalous Residual Surface Conductivity in a Superconductor with Strong Spin-Orbit Coupling}

\author{J. R. Chamorro}
\altaffiliation{These authors contributed equally to this work.}
\affiliation{Department of Chemistry, The Johns Hopkins University, Baltimore, MD 21218, USA}
\affiliation{Institute for Quantum Matter, Department of Physics and Astronomy, The Johns Hopkins University, Baltimore, MD 21218, USA}

\author{P. Chauhan}
\altaffiliation{These authors contributed equally to this work.}
\affiliation{Institute for Quantum Matter, Department of Physics and Astronomy, The Johns Hopkins University, Baltimore, MD 21218, USA}

\author{C. Sun}
\affiliation{Institute for Quantum Matter, Department of Physics and Astronomy, The Johns Hopkins University, Baltimore, MD 21218, USA}

\author{N. Varnava}
\affiliation{Department of Physics \& Astronomy, Rutgers University, Piscataway, NJ 08854, USA}

\author{M. J. Winiarski}
\affiliation{Faculty of Applied Physics and Mathematics and Advanced Materials Center, Gdansk University of Technology, ul. Narutowicza 11/12, 80-233 Gdansk, Poland}

\author{N. Ng}
\affiliation{Department of Chemistry, The Johns Hopkins University, Baltimore, MD 21218, USA}
\affiliation{Institute for Quantum Matter, Department of Physics and Astronomy, The Johns Hopkins University, Baltimore, MD 21218, USA}

\author{H. K. Vivanco}
\affiliation{Department of Chemistry, The Johns Hopkins University, Baltimore, MD 21218, USA}
\affiliation{Institute for Quantum Matter, Department of Physics and Astronomy, The Johns Hopkins University, Baltimore, MD 21218, USA}

\author{L. A. Pressley}
\affiliation{Department of Chemistry, The Johns Hopkins University, Baltimore, MD 21218, USA}
\affiliation{Institute for Quantum Matter, Department of Physics and Astronomy, The Johns Hopkins University, Baltimore, MD 21218, USA}

\author{C. M. Pasco}
\affiliation{Department of Chemistry, The Johns Hopkins University, Baltimore, MD 21218, USA}
\affiliation{Institute for Quantum Matter, Department of Physics and Astronomy, The Johns Hopkins University, Baltimore, MD 21218, USA}

\author{D. Vanderbilt}
\affiliation{Department of Physics \& Astronomy, Rutgers University, Piscataway, NJ 08854, USA}

\author{Yi Li}
\affiliation{Institute for Quantum Matter, Department of Physics and Astronomy, The Johns Hopkins University, Baltimore, MD 21218, USA}

\author{N. P. Armitage}
\affiliation{Institute for Quantum Matter, Department of Physics and Astronomy, The Johns Hopkins University, Baltimore, MD 21218, USA}

\author{T. M. McQueen}
\email{mcqueen@jhu.edu}
\affiliation{Department of Chemistry, The Johns Hopkins University, Baltimore, MD 21218, USA}
\affiliation{Institute for Quantum Matter, Department of Physics and Astronomy, The Johns Hopkins University, Baltimore, MD 21218, USA}
\affiliation{Department of Materials Science and Engineering, The Johns Hopkins University, Baltimore, MD 21218, USA}

\begin{abstract}

\noindent Conventional BCS superconductors are expected to exhibit a conductivity with vanishing dissipation with decreasing temperature. While bulk physical properties measurements indicate PdPb$_{2}$ is a conventional superconductor with a $T_c$ of 3.0 K, measurements of surface impedance through the microwave cavity perturbation technique indicate a large, non-vanishing dissipative component below $T_c$ that is at odds with conventional superconductivity. We demonstrate PdPb$_2$ to be a possible topological superconductor with a fully gapped bulk and a dissipative Majorana fluid surface.
\end{abstract}

\maketitle

\noindent Topological superconductors can host Majorana fermions inside vortex cores and on their surfaces \cite{RevModPhys.82.3045}. They have been proposed as platforms for topological quantum computing \cite{Lian2018, PhysRevLett.86.268}. There are a number of ways of realizing such Majorana fermions, such as topological insulator-superconductor heterostructures \cite{PhysRevLett.114.017001,PhysRevLett.116.257003,He2017,PhysRevLett.112.217001, PhysRevLett.100.096407}, doping-induced superconductivity in topological insulators\cite{Wray2010,PhysRevB.94.180510,PhysRevLett.107.097001,PhysRevLett.107.217001}, and superconductivity-proximitized nanowires\cite{Das2012, Mourik2012, Stanescu2013,PhysRevB.84.144522}. There are also recent proposals of intrinsic topological superconductors, such as in 
$\beta$-Bi$_2$Pd\cite{Li2019,Lv2017,Sakano2015} and FeTe$_{1-x}$Se$_{x}$\cite{Zhang2018,PhysRevLett.117.047001,Machida2019}.

Analogous to a topological insulator, an intrinsic topological superconductor is expected to show surface behavior markedly different from that of the bulk. Topological surface states, if present in the normal state, as is the case in $\beta$-Bi$_2$Pd and FeTe$_{1-x}$Se$_{x}$, may be proximitized by the superconducting bulk, and are expected to possess an unconventional, odd-parity gap function that may give rise to Majorana bound states within vortices. Alternatively, under the right symmetry conditions, the bulk of the superconductor may itself be topological, and can give rise to topological surface states below $T_c$ \cite{PhysRevLett.100.096407}. This kind of topological superconductor is fully gapped in the bulk, but is expected to show non-superconducting surface metallicity. The metallic surfaces in such cases are expected to host Majorana fermions as opposed to massless Dirac fermions, as in topological insulators \cite{PhysRevLett.100.096407}. 

In this Letter, we present evidence for the existence of non-superconducting metallic surface states in centrosymmetric, tetragonal PdPb$_{2}$ below its bulk superconducting transition temperature of $T_{c} = 3$ K. Measurements of the bulk indicate a fully gapped superconducting state, which are in constrast to measurements of the surface impedance through the microwave cavity perturbation technique that indicate significant low frequency dissipation by surface carriers below $T_{c}$, an observation contrary to the expectations for a conventional superconductor. Through extensive sample characterization, we rule out extrinsic origins for the observed phenomenology, and propose PdPb$_{2}$ to be a candidate topological superconductor based on an analysis of the $\mathbb{Z}_2$ invariant of the normal state band structure.

\begin{figure*}[t]
    \includegraphics[width=0.8\textwidth]{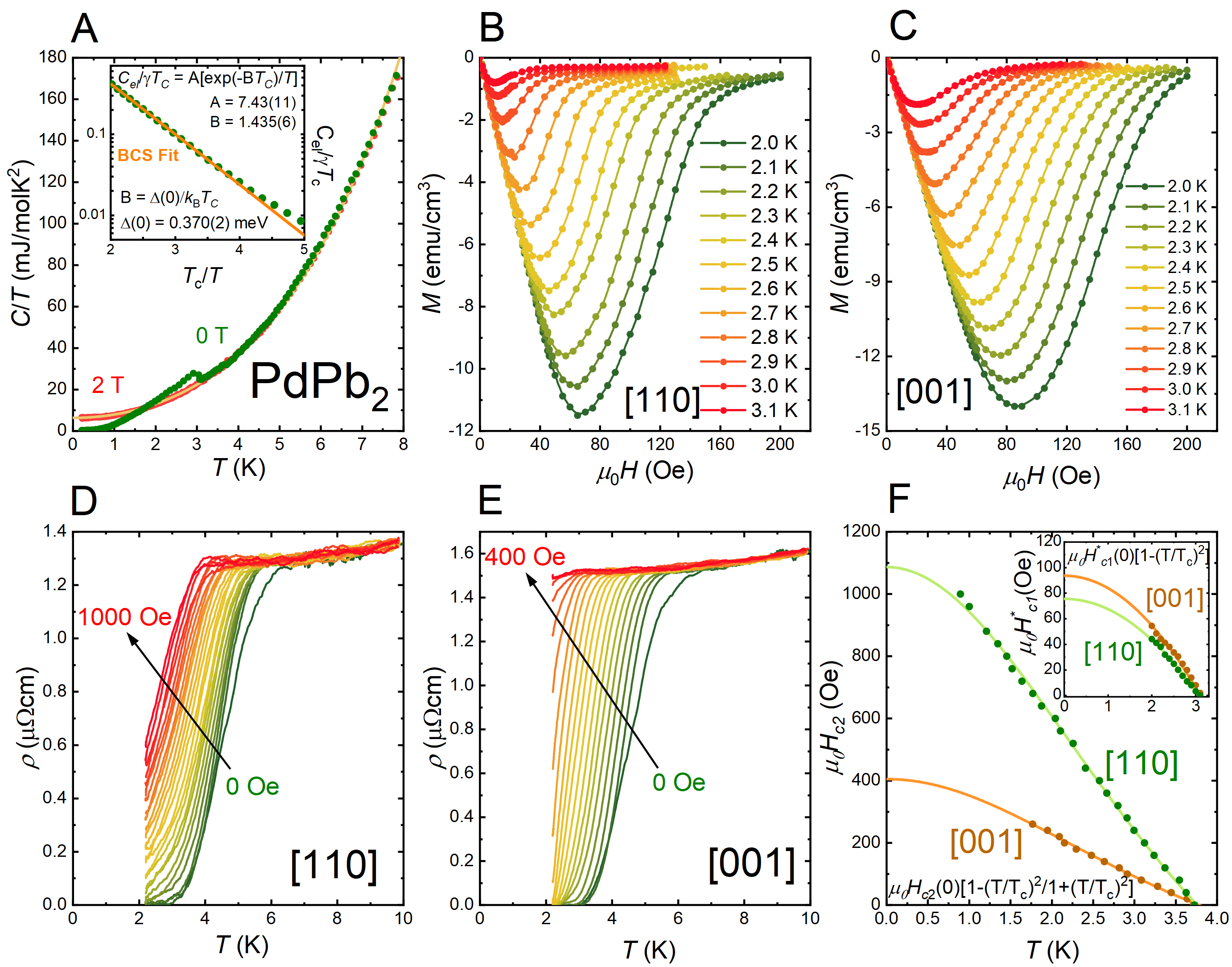}
    \caption{\textbf{A} Heat capacity data of PdPb$_{2}$ at both $\mu_{0}H = 0$ T and 2 T. The 2 T data has been fit to determine the electronic and phonon contributions to the heat capacity in the normal state. The inset shows a BCS fit, described in the text, to the $\mu_{0}H = 0$ T electronic heat capacity of PdPb$_{2}$, which indicates a bulk gap of $\Delta(0) = 0.370(2)$ meV. \textbf{B} and \textbf{C} Field-dependent magnetization curves of PdPb$_{2}$ at various temperatures, with applied fields along the [110] and [001] directions, respectively. \textbf{D} and \textbf{E} Temperature-dependent resistivity curves of PdPb$_{2}$ under various applied fields, with applied fields along the [110] and [001] directions, respectively. \textbf{F} Extracted upper critical fields along both directions, using values determined from the fastest slope of the resistivity curves, shown in \textbf{D} and \textbf{E}. Fits are to the Ginzburg-Landau equation. The inset shows the lower critical fields for both directions, as determined from the magnetization curves after considering demagnetization effects as explained in the Supplementary Material.}
\end{figure*}

Data from measurements of heat capacity, magnetization, and resistivity of a single crystal of PdPb$_{2}$ grown by the Bridgman technique (details in Supplementary Material\cite{SM}) are shown in \textbf{Figure 1}. The heat capacity at zero applied magnetic field, shown in \textbf{Figure 1 A}, demonstrates a superconducting phase transition at $T_{c} = 3.0$ K, as per an equal-entropy approximation, in agreement with previous reports \cite{PhysRevB.38.7067,Havinga1972}. An applied field of $\mu_{0}H = 2$ T is sufficient to quench this transition down to the lowest measured temperature of $T = 0.2$ K. Fits to the $\mu_{0}H = 2$ T data, using $C/T = \gamma + \beta_3 T^2 + \beta_5T^4$, result in an electronic term $\gamma = 6.32(6)$ mJ/mol K$^2$, and two phonon terms $\beta_3 = 1.62(1)$ mJ/mol K$^4$ and $\beta_5 = 0.0184(2)$ mJ/mol K$^6$. The second phonon term arises from the presence of an Einstein mode with $T_E = 49.5(4)$ K, with fits shown in the Supplementary Material \cite{SM}. By subtracting these phonon terms from the $\mu_{0}H = 0$ T data, we are able to obtain the electronic heat capacity of PdPb$_{2}$, $C_{el}$. After normalizing $C_{el}$ by $\gamma$ and $T_{c}$, we may fit the data to $C_{el}/T = A [\mathrm{exp}(-BT_c)/T]$ in order to obtain the magnitude of the superconducting gap $\Delta(0)$, where $A$ and $B$ are two constants \cite{Tari2003}, and $B$ is given by $B = \Delta(0)/k_BT_c$ at temperatures $T_c/T \geq 2$. Our fits, shown in the inset of \textbf{Figure 1 A} on a log scale, indicate a bulk superconducting gap of $\Delta(0) = 0.370(2)$ meV and a $2\Delta(0)/k_BT_c = 2.86$. The origin of this anomalously small value of $2\Delta(0)/k_BT_c$ in relation to the expected BCS value of 3.53 is unknown. The gap value is in agreement with the value of $\Delta(0) \approx 0.4$ meV obtained via microwave cavity measurements, as discussed below and in the supplementary material \cite{SM}. Furthermore, the Sommerfield coefficient $\gamma \rightarrow 0$ as the temperature $T \rightarrow 0 \;\mathrm{K}$, marking the absence of a finite residual electronic contribution to the heat capacity in the bulk, as expected for a fully gapped superconductor.

\begin{table}
\centering
\caption{\label{tab:table1}
Superconducting parameters of a PdPb$_2$ crystal oriented along two crystallographic directions relative to the applied field. These include the lower critical field $H_{c1}^*$, upper critical field $H_{c2}$, thermodynamic critical field $H_c$, coherence length $\xi(T = 0)$, penetration depth $\lambda(T = 0)$, and the Ginzburg-Landau parameter $\kappa$. More information is available in the Supplementary Material \cite{SM}.}
\begin{tabular}{c | c | c}
Parameter & [110] & [001] \\
\hline
\hline
$H_{c1}^*$ (Oe) & 75(6) & 93(4) \\
$H_{c2}$ (Oe) & 1086(26) & 404(17) \\
$H_c$ (Oe) & 191 & 163 \\
$\xi(T = 0)$ (nm) & 55 & 90 \\
$\lambda(T = 0)$ (nm) & 221 & 158 \\
$\kappa$ & 4.02 & 1.75 \\
\end{tabular}
\end{table}

The results of magnetization and resistivity measurements of PdPb$_2$ are shown in \textbf{Figure 1 B-F}. Both measurements show a higher onset $T_c$ into a superconducting state compared to heat capacity and microwave perturbation. It is possible that the higher $T_c$ and breadth of the transition are due to superconducting fluctuations above $T_c$. In order to to rule out extrinsic origins for this behavior, as well as for the surface behavior discussed later in this manuscript, we have performed rigorous measurements of crystal and sample quality, such as powder and single crystal X-ray diffraction, scanning electron microscope imaging, energy-dispersive X-ray spectroscopy, Laue diffraction, and X-ray microtomography, the details of which can be found in the Supplementary Material \cite{SM}. Both X-ray diffraction and energy-dispersive X-ray spectroscopy indicate a nominal stoichiometry of PdPb$_{2}$, with the latter measurement indicating atomic ratios of 33.42(13)\% Pd and $66.58(22)\% \; \mathrm{Pb}$ in the best crystal.

\begin{figure*}[t]
    \includegraphics[width=0.8\textwidth]{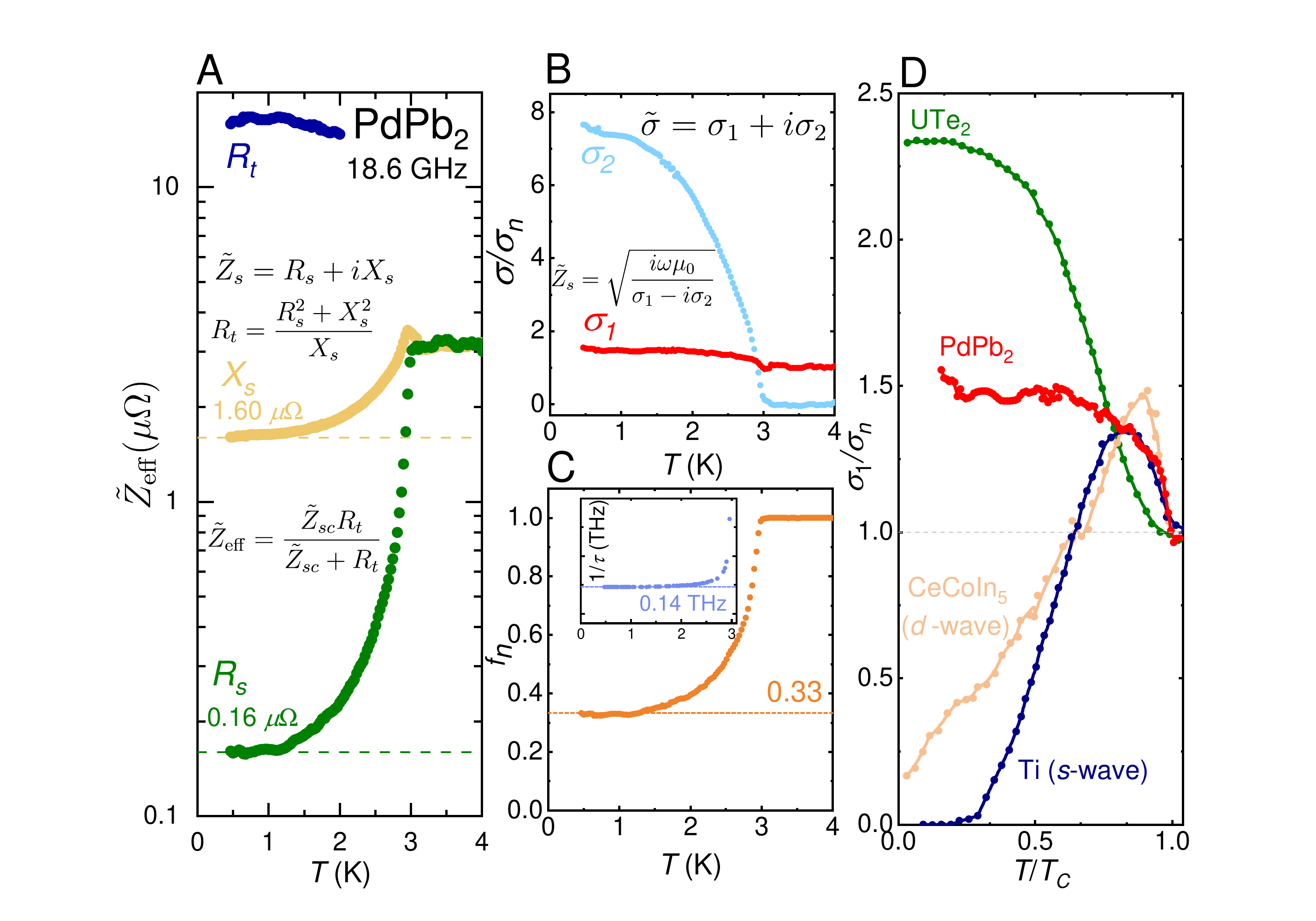}
    \caption{\textbf{A} Temperature-dependent complex impedance $\tilde Z_{\mathrm{eff}}$ of PdPb$_2$. A surface metal sheath with thickness $t$ is considered, giving rise to an effective impedance which arises from both a metal sheath impedance $R_t$ and the impedance of the underlying superconductor $ Z_{sc}$. The real ($R_s$) normal fluid component of $\tilde Z_s$ reaches a finite value of $\sim 0.16 \mu\Omega$, contrary to the expectations for a fully gapped superconductor. \textbf{B} The effective surface complex conductivity of PdPb$_2$ decomposed as given by $\tilde{\sigma} = \sigma_1 - i\sigma_2$, where $\sigma_1$ and $\sigma_2$ are the normal fluid and superfluid responses, respectively. $\sigma_1$ does not go to zero, as expected for a fully gapped superconductor, and instead increases. \textbf{C} The temperature dependence of the fraction of effective normal fluid in PdPb$_2$. This value approaches 33\% instead of zero as $T \rightarrow 0$ K. The inset shows the extracted value of the scattering rate, which plateaus at 0.14 THz at low temperatures. \textbf{D} The real part of the conductivity is shown for several materials, including topological superconductor UTe$_2$ \cite{bae2021}, $s$-wave superconductor Ti \cite{PhysRevB.97.214516}, $d$-wave superconductor CeCoIn$_5$ \cite{Truncik2013}, and PdPb$_2$. Just as in UTe$_2$, $\sigma_1$ anomalously increases with decreasing temperature. The data for Ti, CeCoIn$_5$, and UTe$_2$ were extracted from S. Bae, $et. al.$ \cite{bae2021}}
\end{figure*}

Measurements were performed on two samples, cut from the same Bridgman-grown crystal, with either the [110] or [001] faces exposed, as determined through Laue diffraction. The extracted superconducting parameters are shown in \textbf{Table 1}. The $H_{c2}$ values for both directions, $H_{c2}[110] = 0.1086$ T and $H_{c2}[001] = 0.0404$ T, are much lower than the expected Pauli limit for PdPb$_2$, which through $H_P(\mathrm{T}) \sim 1.85T_c$(K), should be approximately $\mu_0H = 5.5 \; \mathrm{T}$. PdPb$_2$, therefore, does not appear to show signs of triplet pairing or otherwise unconventional superconductivity in the bulk. The derived Ginzburg-Landau parameter $\kappa = \lambda(0)/\xi(0)$, exceeds $\kappa \geq 1/\sqrt{2}$ in both crystallographic directions, indicating PdPb$_2$ is a type-II superconductor.

The microwave cavity perturbation technique was used to determine the temperature-dependent complex surface impedance, $ \tilde{Z}_s(T) $, of a single crystal of PdPb$_2$. Measurements were performed in a superconducting NbTi, $ T_c\approx10 $ K, cylindrical cavity resonator with a resonance frequency of the TE$ _{011} $ mode at 18.66 GHz. The cavity perturbation shifts, marked by a change in complex resonance frequency $ \Delta\tilde{\omega}/\omega_0 $, were measured upon insertion of the sample. For a sample in the skin depth regime, $ \Delta\tilde{\omega}/\omega_0 $ can be related to the impedance $ \tilde{Z}_s = R_s +iX_s $ via $ \frac{\Delta\tilde{\omega}}{\omega_0} = \xi\tilde{Z}_s+\lim_{|\tilde{\sigma}|\to\infty} \frac{\Delta\tilde{\omega}}{\omega_0}$, where $ \xi $ is the resonator constant and $ \lim_{|\tilde{\sigma}|\to\infty} \frac{\Delta\tilde{\omega}}{\omega_0} $ is the metallic shift\cite{SM}. The decomposed surface impedance components, surface resistance $ R_s$ and surface reactance $X_s$, for PdPb$ _2 $ are shown as a function of temperature in \textbf{Figure 2 A}. Both $ R_s $ and $ X_s $ show a sharp change at $ T = 3.0  \; \mathrm{K}$, consistent with a transition into a superconducting state, and reduce with decreasing temperature to plateaued values of $ R_s \sim 0.16 ~\mu\Omega $ and $ X_s \sim 1.60 ~ \mu\Omega $. The non-zero surface resistance of $ \sim 0.16 ~\mu\Omega$ implies the presence of a significant dissipative channel that is odds with conventional electrodynamics.  As discussed below, it may be consistent with a normal surface fluid when the sample is in superconducting state. 

For a bulk homogeneous system, the low temperature dissipation is best quantified via the complex conductivity $ \tilde{\sigma}(T) $, which can be extracted in the local electrodynamics regime, from the complex surface impedance by $ \tilde{Z}_s = \sqrt{i\omega\mu_0/\tilde{\sigma}} $. The decomposed complex conductivity $ \tilde{\sigma} = \sigma_1 - i\sigma_2 $, where $ \sigma_1 $ and $ \sigma_2 $ are the normal fluid and superfluid responses, respectively, is shown in \textbf{Figure 2 B}. For a conventional superconductor, $ \sigma_1 $ is expected to show a Hebel-Schlicter peak near $T_c$ before decreasing with temperature to a minimum, such that $ \sigma_1(0)/\sigma_n = 0$. For d-wave and bulk nodal superconductors, the ratio $ 0 \leq \sigma_1(0)/\sigma_n \leq 1 $ can arise as a result of an underlying nodal gap function. PdPb$_2$, in contrast, shows an anomalous ratio of $ \sigma_1(0)/\sigma_n>1.4 $. Even at the lowest measured temperatures of $T_c/T>3$, the effective value of $ \sigma_1 $ is \textit{greater} than its value in the normal state, indicating that PdPb$_2$ hosts a significant normal dissipative fluid that coexists with fully gapped superconductivity. 

The normal fluid components of the effective complex conductivity for several superconductors are shown in \textbf{Figure 2 D}, as extracted from S. Bae, \textit{et. al.} \cite{bae2021}. These are effective conductivities, since they are calculated as if the system is homogeneous. In the case of elemental Ti, which is an s-wave superconductor with $T_c = 0.39$ K, $\sigma_1$ decreases monotonically to zero as all of $\sigma_n\rightarrow\sigma_2$, i.e. all of the normal ﬂuid becomes superﬂuid below $T_c$. In the case of CeCoIn$_5$, a known d-wave superconductor, the normal ﬂuid component of the complex conductivity decreases to a ﬁnite, non-zero value, as the superconducting gap function possesses nodes. These nodes are sources of dissipation and low energy excitations. Both PdPb$_2$ and topological superconductor candidate UTe$_2$ show an anomalous response, where $ \sigma_1 $ increases with decreasing temperature and subsequently plateaus. In order to determine whether PdPb$ _2 $ possesses non-trivial topology in its electronic band structure, we performed density functional theory (DFT) calculations that are discussed below.  

The anomalous residual low temperature dissipation could arise either from the bulk or the surface of this material. We investigate each of these possibilities in turn. To further explore the anomalous behavior of the complex conductivity in the case of a homogeneous system, we can describe it using a two-fluid model in terms of the normal fluid and superfluid responses, respectively:

\begin{center}
    \centering
	\large{$\tilde{\sigma}(T) = \frac{ne^2}{m^*}\left(\frac{f_s(T)}{i\omega}+ \frac{f_n(T)\tau(T)}{1+i\omega\tau(T)}\right)$}
\end{center}

where $ n $ is the total electron density, $ m^* $ is the effective mass of the electrons, $ \tau$ is the effective normal fluid scattering lifetime, $ f_n $ is the normal fluid fraction and $ f_s = 1-f_n $ is the superfluid fraction. As $\tau$ is always a real positive number, the normal fluid fraction must satisfy $ f_n\ge2/(1+\sqrt{1+(\sigma_2/\sigma_1)^2}) $. For conventional superconductors, as $ T\rightarrow0 $ K, the ratio $ \sigma_2/\sigma_1\rightarrow\infty $, implying $ f_n\rightarrow0 $. As shown in \textbf{Figure 2 B}, this ratio approaches $ \sigma_2/\sigma_1\rightarrow 5 $ in PdPb$ _2 $ as $ T\rightarrow0 $, giving a minimum residual normal fluid fraction $ f_n \ge33\% $, i.e. one-third of the normal state value, as shown in \textbf{Figure 2 C}. This result is similar to one observed recently in the topological superconductor candidate UTe$_2 $ \cite{bae2021}. Again, this is an effective density calculated from a model for the conductivity that assumes that system has a homogeneous electrodynamic response. Similar to UTe$_2$, PdPb$_2$ possesses signiﬁcant dissipation on the surface against the backdrop of a fully gapped bulk superconducting state. Furthermore, as shown in the inset of \textbf{Figure 2 C}, we determine from the above equation the inverse scattering rate for the anomalous fluid, which decreases to a plateau at low temperature to a value of $ \tau^{-1}\approx0.14 $ THz. This value is too low to create an impurity-induced normal fluid scenario. 

A possible origin for the anomalous effective conductivity and significant dissipative response is a surface origin of a dissipative normal conduction channel. If the surface of the bulk superconductor has a normal conductor sheath then the expression for extracting a bulk conductivity from the surface impedance becomes invalid.  The extracted conductivity can naturally show anomalous features due to contributions from the normal conduction channel.  One would need to account for such a thin normal fluid layer at the surface in the effective impedance of the sample. For a thin normal conductor layer with 2D resistance $ R_t $ on top of a bulk superconductor with impedance $ Z_{sc} $ the effective impedance of the system is given to good approximation by $ Z_\text{{eff}} = Z_{sc}R_t/(Z_{sc}+R_t) $\cite{SM}.  We assume that the bulk superconductor has a purely imaginary impedance $ Z_{sc} = iX_{sc} $(reasonable for a fully gapped 3D superconductor at low temperatures).  Then normal channel's response follows from the measured signal as $ R_t = \frac{R_s^2 + X_s^2}{X_s} $.  \textbf{Figure 2 A} shows the impedance $ R_t $ of the thin normal layer with $ R_t \sim 16~\mu\Omega$ for $ T\rightarrow0 $ K. $ R_t $ reduces monotonically with increasing temperature. One can see that it is reasonably temperature independent in the low temperature region where this analysis applies.

\begin{figure*}[t]
    \includegraphics[width=1\textwidth]{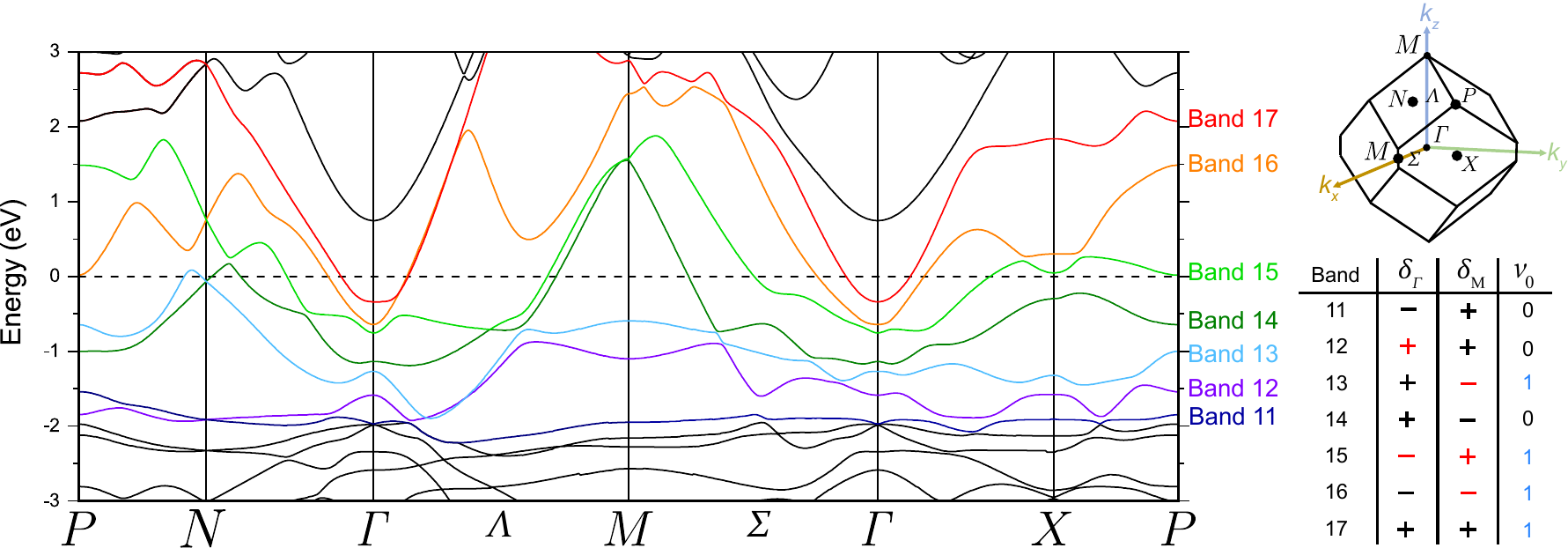}
    \caption{The calculated electronic band structure of PdPb$_2$, calculated using the local density approximation and with spin-orbit coupling considered. The first Brillouin zone for body-centered tetragonal is shown, under the $a > c$ lattice parameter condition. The parities at $\Gamma (\delta_\Gamma)$ and $M (\delta_M)$ are shown as positive and negative signs. A change in parity from one band to the next is shown in red. The $\mathbb{Z}_2$ topological invariant $\nu_0$ is shown for each band, which is calculated as the product of the parities at $\Gamma$ and $M$ for that band, along with the product of those two parities for all previous bands. Non-trivial invariants of $\nu_0 = 1$ are shown in blue.}
\end{figure*}

The DFT calculated electronic band structure, with spin-orbit coupling included, is shown in \textbf{Figure 3}. PdPb$_2$ has been predicted to be a symmetry-enforced semimetal with line crossings between the two high symmetry points of the first Brillouin zone $\Gamma$ and $M$ \cite{Bradlyn2017,Vergniory2019}. It has also been predicted to be a topological insulator based on the absence of a linear combination (NLC) of the elementary band representations (EBRs) across the Brillouin zone \cite{Bradlyn2017}, though it is not an insulator. Our calculated band structure, as well as those previously calculated, indeed show band crossings along the $\Lambda$ path along $k_z$ between $\Gamma$ and $M$, with point group $4mm$. Since PdPb$_2$ possesses time-reversal symmetry, its bulk topology is characterized by a $\mathbb{Z}_2$ invariant \cite{PhysRevLett.98.106803, PhysRevB.75.121306, Roy2010}. As it further possesses inversion symmetry, the $\mathbb{Z}_2$ invariant can be calculated readily by considering the product of the parities of all of the bands up to and including the ones that cross the Fermi level at both $\Gamma$ and $M$ \cite{PhysRevB.76.045302}. Only these two special points in the Brillouin zone determine the topological invariant for body-centered tetragonal, as of the eight time-reversal invariant momentum (TRIM) points in three-dimensions, $\Gamma$ and $M$ are the only two with odd coefficients ($1\Gamma,4N,2X,1M$). The band parities at $\Gamma$ and $M$, and the resulting topological invariant $\nu_0$, which also considers the product of the parities of all bands beneath the one in question, are shown in \textbf{Figure 3}. Our calculations indicate that four of the five bands that cross the Fermi level have non-trivial invariants of $\nu_0 = 1$. PdPb$_2$ is therefore a $\mathbb{Z}_2$ metal in the normal state, a designation also used to describe other topological superconductor candidates such as FeTe$_{1-x}$Se$_{x}$\cite{Zhang2018,PhysRevLett.117.047001,Machida2019}, $\beta$-Bi$_2$Pd\cite{Li2019,Lv2017,Sakano2015}, doped Bi$_2$Se$_3$ \cite{PhysRevLett.107.217001, Liu2015}, Tl$_5$Te$_3$ \cite{PhysRevLett.112.017002}, and more recently the AV$_3$Sb$_5$ (A = K, Rb, and Cs) family of materials \cite{PhysRevMaterials.3.094407,PhysRevMaterials.5.034801,PhysRevLett.125.247002,Yang2020}. More in depth theoretical studies are needed to disentangle the influence of the topological band crossings near the Fermi level on the topology and physical properties.

While our calculations do not reveal the presence of surface states near the Fermi level in the normal state (not shown), the non-trivial invariant of bands that cross the Fermi level in PdPb$_2$ could result in a topological superconductor \cite{PhysRevLett.105.097001}. Strong spin-orbit coupling arising from the heavy elements Pd and Pb can drive the system into a topological ground state characterized by a fully gapped bulk superconducting state with topologically non-trivial Majorana surface states. We believe that low temperature microwave dissipation is a generic signature of such surface states. PdPb$_2$ therefore adds on to a short list of three-dimensional topological superconductor candidates. 

In conclusion, we have discovered an anomalous, large, dissipative surface electrodynamic response in the centrosymmetric superconductor PdPb$_2$.  A natural explanation of this phenomenon is that this material is a topological superconductor with surface Majoranas states.  The ease with which large, high quality single crystals can be grown can make this material ideal for understanding the origin of this anomalous behavior.

\section*{Acknowledgments}

This work was supported as part of the Institute for Quantum Matter, an Energy Frontier Research Center funded by the U.S. Department of Energy, Office of Science, Office of Basic Energy Sciences, under Award DE-SC0019331. J. R. C. acknowledges support of the Gompf Family Fellowship from the JHU Department of Chemistry. J. R. C. is grateful to M. A. Siegler for assistance collecting single-crystal X-ray diffraction data. This work utilized synthetic capabilities in the Platform for the Accelerated Realization, Analysis, and Discovery of Interface Materials (NSF DMR-1539918), a National Science Foundation Materials Innovation Platform. The Magnetic Properties Measurement System used for data collection  was  funded  by  the National Science Foundation,  Division of Materials Research, Major Research Instrumentation Program, under Award No. 1828490. The dilution refrigerator was funded by the National Science Foundation, Division of Materials Research, Award No. 0821005. Electronic band structure calculations were conducted using computational resources at the Maryland Advanced Research Computing Center (MARCC).

\bibliography{biblio}

\end{document}